# Learning Common Harmonic Waves on Stiefel Manifold – A New Mathematical Approach for Brain Network Analyses

*Jiazhou Chen[1,2], Guoqiang Han[1], Hongmin Cai[1], Defu Yang[2,3], Paul J. Laurient[4], Martin Styner[2,5], Guorong Wu[2,5], and Alzheimer's Disease Neuroimaging Initiative (ADNI)*

[1] *School of Computer Science and Engineering, South China University of Technology, Guangzhou, China.*
[2] *Department of Psychiatry, University of North Carolina at Chapel Hill, USA.*
[3] *Intelligent Information Processing Laboratory, Hangzhou Dianzi University, Hangzhou, China.*
[4] *Department of Radiology, Wake Forest School of Medicine, USA.*
[5] *Department of Computer Science, University of North Carolina at Chapel Hill, USA.*

## Abstract

Converging evidence shows that disease-relevant brain alterations do not appear in random brain locations, instead, its spatial pattern follows large-scale brain networks. In this context, a powerful network analysis approach with a mathematical foundation is indispensable to understand the mechanism of neuro-pathological events spreading throughout the brain. Indeed, the topology of each brain network is governed by its native harmonic waves, which are a set of orthogonal bases derived from the Eigen-system of the underlying Laplacian matrix. To that end, we propose a novel connectome harmonic analysis framework to provide enhanced mathematical insights by detecting frequency-based alterations relevant to brain disorders. The backbone of our framework is a novel manifold algebra appropriate for inference across harmonic waves that overcomes the limitations of using classic Euclidean operations on irregular data structures. The individual harmonic difference is measured by a set of common harmonic waves learned from a population of individual Eigen-systems, where each native Eigen-system is regarded as a sample drawn from the Stiefel manifold. Specifically, a manifold optimization scheme is tailored to find the common harmonic waves which reside at the center of Stiefel manifold. To that end, the common harmonic waves constitute the new neurobiological bases to understand disease progression. Each harmonic wave exhibits a unique propagation pattern of neuro-pathological burdens spreading across brain networks. The statistical power of our novel connectome harmonic analysis approach is evaluated by identifying frequency-based alterations relevant to Alzheimer's disease, where our learning-based manifold approach discovers more significant and reproducible network dysfunction patterns compared to Euclidian methods.

# 1. Introduction

Recent advances in neuroimaging offer an in-vivo and non-invasive window for investigating connectivity between brain regions [1-3]. For example, the combination of diffusion-weighted magnetic resonance imaging (DW-MRI) and tractography technology can be used to reconstruct major fiber bundles in the brain allowing for the visualization of the structural pathways that connect distant brain regions [4]. The ensemble of macroscopic brain connections can then be described as a complex network - the 'connectome'. Various computational and statistical inference methods have been developed to characterize diverse properties of complex networks and then identify network differences in terms of nodes, links, or even subgraphs that are associated with neurological disorders [1, 3]. Due to the high dimensionality of the brain connectome, it is a common practice to analyze node-wise graph metrics such as local clustering coefficient and small-worldness [5], instead of using whole-brain connectivity information. By doing so, however, it becomes difficult to discover topological patterns which are an essential aspect of network analyses. On the flip side, there are also a plethora of methods proposed to quantify network changes at the level of individual links rather than nodes [6-12]. Like node-wise analyses, link-wise analyses are univariate in nature and disregard the multivariate network structure. In addition, due to high dimensionality, link-wise significance tests necessitate strict multiple-comparison correction to alleviate the issue of false positives, which potentially discards scientifically meaningful links [13].

Many neuroimaging studies have found that the progression of neuropathology follows the topology of large-scale networks in the brain [1, 14-17]. For instance, a network diffusion model was used in [14, 15] to predict the disease progression in dementia, where the diffusion process is governed by the Laplacian matrix of the underlying brain network. Like various natural phenomena, the Eigen-system of the Laplacian also constitutes the basis of self-organizing patterns (shown in the bottom of **Fig. 1**), where each specific harmonic wave is indeed the Eigen-vector associated with a particular frequency (Eigen-value). Harmonic-based analyses have been used to investigate frequency-based alterations in neuropsychiatric diseases [18, 19] and functional neural activity [20]. Since the harmonic waves are orthogonal to each other, encoding brain connectivity via the harmonic domain offers great flexibility for the performance of group difference analyses.

However, current harmonic analysis approaches have two major limitations. (1) Lack of an unbiased reference to measure the difference between individuals. In general, an unbiased reference domain is necessary for conducting group comparisons to provide standardized measurements for the statistical analyses. For example, since intrinsic structural differences are often mixed with external differences (such as the size and shape of the brain), an atlas image is used as a standard spatial reference for voxel-based morphometry (VBM) [21]. The morphometry differences of interest (such as gray matter density [22]) among the spatially normalized images are thought to be more relevant to neurobiological processes. Yet, different networks lead to various Eigen-systems, and thus a harmonic reference space for brain networks needs to represent the common set of harmonic waves that can appropriately represent the majority of the individual Eigen-systems. (2) Lack of the appropriate manifold algebra. Despite the well-known importance of a reference space in neuroimaging, finding such reference spaces for manifold data, such as brain networks, it is still an open problem as the complexity of data geometry (topology) is beyond regular data arrays [23, 24]. As shown in the top of **Fig. 1**, current approaches treat high-dimensional network data as a regular matrix or vector. Although applying Euclidean algebra to average brain networks [25] or diffuse connectivity information [26] on a link-wise basis is straightforward, the resulting group-mean network may no longer contain the essential network topology disrupting the geometry of Eigen-systems.

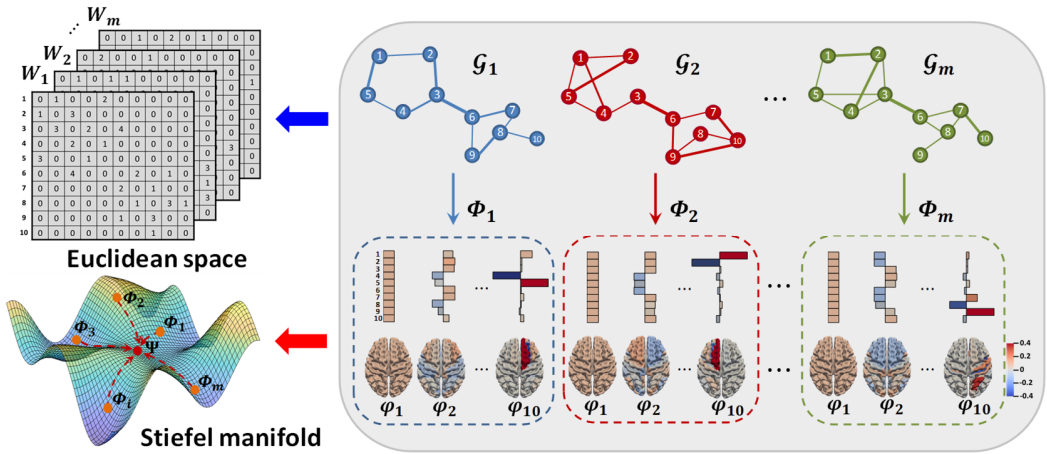

**Fig. 1** Conventional methods (top) apply the classic Euclidean operations on the graph structure. Such rigid operations underestimate the irregular data structures and yield poor performances. In comparison, our learning-based approach (bottom) fully respects the irregular graph data structure and discovers brain network harmonics on the Stiefel manifold.

To overcome these limitations, we propose a novel manifold learning method to discover the unbiased population mean of individual Eigen-systems. Since each Eigen-vector is orthogonal to all others, it is reasonable to assume that each Eigen-system behind the individual brain network resides on a high-dimensional Stiefel manifold [27]. Since each Eigen-system is uniquely associated with the corresponding underlying propagation patterns in the brain network, the well-studied Stiefel manifold (red arrow in **Fig. 1**) allows us to find a set of common bases that appropriately express the network propagation patterns, as compared to Euclidean algebra which lacks well-defined algebraic operators for manifold data.

Specifically, our method iteratively alternates two steps. (1) Adjust each native Eigen-system toward the latent manifold mean. The construction of each Eigen-system is not only influenced by the underlying Laplacian matrix but also attracted by the latent common harmonic waves at the manifold center. (2) Update the manifold center. We first project each Eigen-system to the tangent space at the current manifold center. Then, we estimate the mean tangent which points to the new location of the manifold mean. After mapping the mean tangent back to the Stiefel manifold, we can obtain the new estimation of the manifold center that is used to guide the refinement of individual Eigen-systems in Step 1. The outcome of our manifold optimization is a set of orthogonal vectors located at the manifold center, which represent the common harmonic waves learned over the population of brain networks.

As each harmonic wave exhibits a unique self-organized oscillation pattern across the brain network, our learned set of common harmonic waves offers a new window to investigate the mechanism of neurodevelopment or neurodegeneration in the setting of brain networks using the classic physics concepts such as power and energy [19]. We have evaluated the statistical power of our new network harmonic analysis approach in a brain network study of Alzheimer's disease (AD). Compared to the conventional approach [25] using Euclidean operations, more statistically significant and reliable frequency-based alterations have been discovered using the common harmonic waves learned on Stiefel manifold.

## 2. Method

First, we provide the brief background on spectral graph theory and Stiefel manifold optimization in Section 2.1. The motivation for discovering common harmonic waves for brain network analyses is explained in Section 2.2. Then we present our manifold learning method for common harmonic waves in Section 2.3, followed by the optimization scheme in Section 2.4. The application of the learned common harmonic waves using a neuroimaging dataset is demonstrated in Section 2.5. The notation used in this paper is summarized in **Table I** for ease of exposition.

Table I List of notations used in this paper

| Notation | Remark |
|---|---|
| $x, \boldsymbol{x}, \boldsymbol{X}$ | Scalar, vector and matrix |
| $\mathcal{G}(V, \mathcal{E}, \boldsymbol{W})$ | A graph $\mathcal{G}$ with nodes $V$, edges $\mathcal{E}$ and weights $\boldsymbol{W}$ |
| $\boldsymbol{L}$ | A Laplacian matrix of graph $\mathcal{G}$ |
| $\boldsymbol{\Phi}_s$ | $s^{th}$ individual network harmonic waves |
| $\boldsymbol{\Psi}$ | Common network harmonic waves |
| $\mathbb{R}^n$ | $n$-dimensional real space |
| $\mathcal{O}^n$ | Orthogonal group consisting of $n$-by-$n$ orthogonal matrices |
| $\mathcal{M}_H$ | The Stiefel manifold of harmonic waves |
| $\mathcal{V}(n,p)$ | $(n,p)$-Stiefel manifold |
| $\mathcal{T}_X, \Delta$ | Tangent space and tangent vector of manifold at $\boldsymbol{X}$ |
| $exp$ | Exponential map |
| $F_X$ | Matrix derivative of some function $F$ with respect to $\boldsymbol{X}$ |
| $\nabla_X F$ | Gradient of $F$ at point $\boldsymbol{X}$ in manifold space |

## 2.1 Background

**Graph spectrum and harmonic waves.** Each brain network can be encoded in a graph $\mathcal{G} = (V, \mathcal{E}, \boldsymbol{W})$, where $V = \{v_i | i \in 1, \cdots, n\}$ represents the node set with $n$ nodes and $\mathcal{E} = \{e_{ij} | (v_i, v_j) \in V \times V\}$ is the set of all possible links. Let $\boldsymbol{W} \in \mathbb{R}^{n \times n}$ be a weighted adjacency matrix where each element $w_{ij}$ in $\boldsymbol{W}$ measures the connectivity strength between node $v_i$ and $v_j$. Then the symmetric graph Laplacian matrix $\boldsymbol{L}$ of the underlying graph can be calculated by:

$$\boldsymbol{L} = \boldsymbol{D} - \boldsymbol{W} \tag{1}$$

where $\boldsymbol{D} = diag(d_1, d_2, \ldots, d_n)$ is the degree matrix of the graph. Each diagonal element equals to the total connectivity degree of the underlying node, i.e., $d_i = \sum_{j=1}^{n} w_{ij}$.

A set of harmonic waves $\boldsymbol{\Phi}$ can be obtained by:

$$\min_{\boldsymbol{\Phi} \in \mathbb{R}^{n \times n}} tr(\boldsymbol{\Phi}^T \boldsymbol{L} \boldsymbol{\Phi}), \quad s.t. \quad \boldsymbol{\Phi}^T \boldsymbol{\Phi} = \boldsymbol{I}_p \tag{2}$$

where $tr(\cdot)$ is the trace operator and $\boldsymbol{I}_p \in \mathbb{R}^{p \times p}$ stands for the identity matrix. The optimization problem in Eq. (2) has the deterministic solution $\widehat{\boldsymbol{\Phi}}$, which is the set of Eigen-vectors of the matrix $\boldsymbol{L}$. Without loss of generality, we can sort each Eigen-vector in $\widehat{\boldsymbol{\Phi}}$, column by column, in increasing order of Eigen-values. Given the connected graph $\mathcal{G}$ (i.e., no isolated nodes), the first smallest Eigen-value is always zero and each element in the associated Eigen-vector is a constant. As the Eigen-value increases, the corresponding Eigen-vector exhibits more and high frequency patterns (more rapid and localized oscillations) across the brain network, as shown in the bottom of **Fig. 1**.

**The Stiefel manifold** is a well-studied space and is defined as a set of ordered orthonormal $p$-frames of vectors in $\mathbb{R}^n$, denoted by $\mathcal{V}(n,p)$. In this context, any matrix $X \in \mathbb{R}^{n \times p}$ can be regarded as a point sitting on Stiefel manifold $\mathcal{V}(n,p)$ as long as $X$ is an orthogonal matrix, i.e., $X^T X = I_p$. A special case is when $p = 1$, the Stiefel manifold reduces to the set of all unit vectors, which forms the unit sphere. Another special case is when $p = n$, the Stiefel manifold is the group of orthogonal $n \times n$ matrices $\mathcal{V}(n,n) \in \mathcal{O}^n$. Since there is no analytical formula for endpoint geodesics on the Stiefel manifold (*i.e.*, locally shortest length curves between two points $X \in \mathcal{V}(n,p)$ and $Y \in \mathcal{V}(n,p)$), it is common to approximate the geodesic between $X$ and $Y$ in the ambient space by the following squared distance [27]:

$$d^2(X,Y) = \frac{1}{2} tr(X-Y)^T(X-Y) = p - tr(X^T Y) \tag{3}$$

**Gradient descent optimization on Stiefel manifold.** For a point $X \in \mathcal{V}(n,p)$, the tangent space $\mathcal{T}_X$ at $X$ consists of a set of tangents $\{\Delta\}$ such that $X^T \Delta = 0$. Suppose a real-valued function $F$ is smooth on the Stiefel manifold, the gradient of function $F$ at $X$, *i.e.*, $\nabla_X F \in \mathcal{T}_X$, can be obtained by [28]:

$$\nabla_X F = F_X - X F_X^T X \tag{4}$$

where $F_X$ stands for the matrix derivative of function $F$ with respect to $X$. The gradient calculation in Eq. 4 plays an important role in the application of Stiefel manifold optimization such as nonlinear mean shift [29]. The efficient calculation of the manifold gradient on the flattened tangent space offers a reasonable descent direction for optimizing function $F$ on the manifold. After that, an exponential mapping operation is required to map a tangent $\Delta \in \mathcal{T}_X$ back onto the Stiefel manifold by:

$$exp_X(\Delta) = XB + QC \tag{5}$$

where the matrices $B$, $Q$, and $C$ are calculated in two steps. (1) Apply compact QR decomposition of $(I - XX^T)\Delta$ [30] and thus obtain matrices $Q \in \mathbb{R}^{n \times p}$ and $R \in \mathbb{R}^{p \times p}$. (2) Solve $B \in \mathbb{R}^{p \times p}$ and $C \in \mathbb{R}^{p \times p}$ by:

$$\begin{bmatrix} B \\ C \end{bmatrix} = \exp\left(\begin{bmatrix} A & -R^T \\ R & 0_p \end{bmatrix}\right) \begin{bmatrix} I_p \\ 0_p \end{bmatrix} \tag{6}$$

where $A = X^T \Delta$ is a $p \times p$ matrix and $0_p \in \mathbb{R}^{p \times p}$ stands for the zero matrix.

## 2.2 Problem Statement

For each brain network $\mathcal{G}_s$ ($s = 1, \ldots, m$), we can obtain its set of harmonic waves $\mathbf{\Phi}_s$ (an orthogonal matrix) by applying Eigen-decomposition on its Laplacian matrix $\mathbf{L}_s$. We are interested in finding the mean (denoted by $\mathbf{\Psi}$) of $m$ individual harmonic waves $\{\mathbf{\Phi}_s | s = 1, \ldots, m\}$, where $\mathbf{\Psi}$ is called the common harmonic waves which are also an orthogonal matrix.

Although it is efficient to calculate $\mathbf{\Psi}$ by simple averaging, i.e., $\overline{\mathbf{\Psi}} = \frac{1}{m}\sum_{s=1}^{m}\mathbf{\Phi}_s$, the column vectors in $\overline{\mathbf{\Psi}}$ are not orthogonal to each other any longer, which compromises the applicability of $\overline{\mathbf{\Psi}}$ as the meaningful harmonic waves. Considering that each harmonic set $\mathbf{\Phi}_s$ resides on the high dimensional Stiefel manifold $\mathcal{M}_H$, the common harmonic waves $\overline{\mathbf{\Psi}}$ generated by arithmetic averaging results in a $\overline{\mathbf{\Psi}}$ may not be directly located on the same manifold as all the individual harmonic waves.

An alternative way is to average over the adjacency matrices $\{\mathbf{W}_s | s = 1, \ldots, m\}$ and then calculate the Eigen-system $\widetilde{\mathbf{\Psi}}$ based on the Laplacian matrix which is derived from the average of adjacency matrices $\overline{\mathbf{W}} = \frac{1}{m}\sum_{s=1}^{m}\mathbf{W}_s$. However, such Euclidean operations are highly sensitive to noises and/or outlying data points. In addition, a heuristic assumption that the intrinsic complex geometry of brain network and Eigen-system data can be well expressed in Euclidean space is difficult to satisfy.

Given that each harmonic set $\mathbf{\Phi}_s$ is an orthogonal matrix, it is reasonable to consider finding the latent common harmonic waves $\mathbf{\Psi}$ on the Stiefel manifold. It is worth noting that the graph spectrum of each brain network is spanned by its harmonic waves $\mathbf{\Phi}_s$, sorted from low frequency to high frequency [23]. Since the harmonic waves associated with high frequency (larger eigenvalues) are more sensitive to possible noise, we only consider the first $p$ ($p \leq n$) harmonic waves in each $\mathbf{\Phi}_s$. In the following, we regard $\mathbf{\Phi}_s \in \mathcal{M}_H$ ($\mathcal{M}_H \subset \mathcal{V}(n,p)$) as an $n \times p$ orthogonal matrix unless otherwise stated.

## 2.3 Learning Common Harmonic Waves on Stiefel Manifold

Given $m$ Laplacian matrices $\{\mathbf{L}_1, \mathbf{L}_2, \ldots, \mathbf{L}_m\}$, we simultaneously estimate the native harmonic waves $\mathbf{\Phi}_s$ for each $\mathbf{L}_s$ and optimize the common harmonic waves $\mathbf{\Psi}$, which are both optimized on the Stiefel manifold $\mathcal{M}_H$. Specifically, we require the latent common harmonic waves to be located at the manifold center that has the shortest geodesic distances to all individual harmonic waves $\{\Phi_s\}$. To that end, we opt to minimize $\sum_{s=1}^{m} d^2(\Phi_s, \Psi)$. By integrating Eq. 2 and Eq. 3, the objective function becomes:

$$\min_{\{\Phi_s\},\Psi} \sum_{s=1}^{m} \{tr(\Phi_s^T L_s \Phi_s) + \lambda(p - tr(\Phi_s^T \Psi))\} \quad (7)$$

$$s.t. \ \forall s: \Phi_s^T \Phi_s = I_p$$

where $\lambda$ is a scalar balancing the self-independence of estimating each $\Phi_s$ and the joint-collaboration of optimizing the latent common harmonic waves $\Psi$.

Since it is computationally expensive to estimate $\{\Phi_s\}$ and $\Psi$ jointly, we propose the following gradient descent manifold optimization under the framework of ADMM (Alternating Direction Method of Multipliers) [31], where the augmented Lagrange function becomes:

$$\arg\min_{\{\Phi_s\},\Psi} \sum_{s=1}^{m} F_{\Phi_s,\Psi} =$$

$$\arg\min_{\{\Phi_s\},\Psi} \sum_{s=1}^{m} \{tr(\Phi_s^T L_s \Phi_s) + \lambda(p - tr(\Phi_s^T \Psi)) + tr(\Lambda_s^T(\Phi_s^T \Phi_s - I_p))\}, \quad (8)$$

where each $\Lambda_s$ ($s = 1, \dots, m$) is a $p \times p$ factor matrix of the Lagrange multipliers.

**2.4 Optimization Scheme**

We optimize the objective function in Eq. 8 in two alternative steps.

**Sub-problem 1: Estimating each native harmonic set $\Phi_s$.** Since the harmonic waves $\{\Phi_s\}$ are independent, we can estimate each $\Phi_s$ separately by fixing $\Psi$. By dropping the unrelated variables, the objective function becomes:

$$\arg\min_{\Phi_s} F_{\Phi_s} = \arg\min_{\Phi_s}\{tr(\Phi_s^T L_s \Phi_s - \lambda \Phi_s^T \Psi) + tr(\Lambda_s^T(\Phi_s^T \Phi_s - I_p))\} \quad (9)$$

It is worth noting that the individual harmonic waves $\Phi_s$ are not only determined by its own Laplacian matrix $L_s$, but also influenced by the latent common harmonic waves $\Psi$. Since Eq. 9 is a typical quadratic problem on the Stiefel manifold, it is often required that $L_s$ is positive definite. Therefore, we first replace $L_s$ with $\tilde{L}_s = \beta I - L_s$, where a relaxation parameter $\beta$ is used to ensure $\tilde{L}_s$ is a positive definite matrix. We set $\beta$ as the greatest eigenvalue of $\boldsymbol{L}_s$. By doing so, the minimization of Eq. 9 becomes the maximizing:

$$\arg\max_{\boldsymbol{\Phi}_s} \tilde{F}_{\boldsymbol{\Phi}_s} = \arg\max_{\boldsymbol{\Phi}_s} \{tr(\boldsymbol{\Phi}_s^T \tilde{\boldsymbol{L}}_s \boldsymbol{\Phi}_s) + \lambda tr(\boldsymbol{\Phi}_s^T \boldsymbol{\Psi}) - tr(\boldsymbol{\Lambda}_s^T(\boldsymbol{\Phi}_s^T \boldsymbol{\Phi}_s - \boldsymbol{I}_p))\} \quad (10)$$

The last term in Eq. 10 is the Lagrange multiplier. We can solve $\boldsymbol{\Phi}_s$ via the KKT condition as:

$$\frac{\partial \tilde{F}_{\boldsymbol{\Phi}_s}}{\partial \boldsymbol{\Phi}_s} = 2\tilde{\boldsymbol{L}}_s \boldsymbol{\Phi}_s + \lambda \boldsymbol{\Psi} - 2\boldsymbol{\Phi}_s \boldsymbol{\Lambda}_s = 0 \quad (11)$$

To overcome the instability as well as reduce the computational cost caused by the matrix inversion involved in $\tilde{\boldsymbol{L}}_s$ and $\boldsymbol{\Lambda}_s$, we adopt the generalized power iteration (GPI) from [32] to iteratively optimize $\boldsymbol{\Phi}_s$ in the following four steps:

(1) Initialize $\boldsymbol{\Phi}_s$ as the Eigen-vector matrix after applying SVD to the underlying Laplacian matrix $\boldsymbol{L}_s$.
(2) Update $\boldsymbol{\Theta} \leftarrow \tilde{\boldsymbol{L}}_s \boldsymbol{\Phi}_s + \lambda \boldsymbol{\Psi}$.
(3) Calculate $\boldsymbol{\Phi}_s$ by maximizing $tr(\boldsymbol{\Phi}_s^T \boldsymbol{\Theta})$ and subject it to the orthogonal constraint $\boldsymbol{\Phi}_s^T \boldsymbol{\Phi}_s = \boldsymbol{I}_p$. We can derive the closed-form solution by $\boldsymbol{\Phi}_s = \boldsymbol{U}\boldsymbol{V}^T$, where $\boldsymbol{U}$ and $\boldsymbol{V}$ are the left and right Eigen matrix after the full SVD on $\boldsymbol{\Theta}$. (please refer to [32] for detail)
(4) Iteratively perform the steps (2)-(3) until convergence.

**Sub-problem 2: Estimating the common harmonic set $\boldsymbol{\Psi}$.** Given the individual harmonic waves $\{\boldsymbol{\Phi}_s\}$, the objective function of $\boldsymbol{\Psi}$ is:

$$\arg\min_{\boldsymbol{\Psi}} \sum_{s=1}^{m} d^2(\boldsymbol{\Phi}_s, \boldsymbol{\Psi}) = \arg\min_{\boldsymbol{\Psi}} \sum_{s=1}^{m} (p - tr(\boldsymbol{\Phi}_s^T \boldsymbol{\Psi})) \quad (12)$$

The intuition behind in Eq. 12 is to find the latent mean $\boldsymbol{\Psi}$ on the Stiefel manifold which has the shortest geodesic distances to all the observed samples $\{\boldsymbol{\Phi}_s\}$ residing on the Stiefel manifold. Thus, our optimization falls into the classic problem of solving the Fréchet mean on the Stiefel manifold which can be efficiently solved by adopting the Weiszfeld algorithm [33]. Specifically, we alternately perform the following two steps until convergence:

(1) Suppose $\boldsymbol{\Psi}^{(k)}$ is the current estimation of the manifold center. We calculate the gradient $\nabla_{\boldsymbol{\Psi}}$ of the energy function in Eq. 12 with respect to each $\boldsymbol{\Phi}_s$ as: $\nabla_{\boldsymbol{\Psi}} d^2(\boldsymbol{\Psi}^{(k)}, \boldsymbol{\Phi}_s) = \boldsymbol{\Psi}^{(k)} \boldsymbol{\Phi}_s^T \boldsymbol{\Psi}^{(k)} - \boldsymbol{\Phi}_s$, which are denoted by the black arrows in **Fig. 2**. Then, the mean tangent $\Delta\Psi^{(k+1)} \in \mathcal{T}_\Psi$ can be efficiently obtained by:

$$\Delta\Psi^{(k+1)} = -\sum_{s=1}^{m} \nabla_\Psi d^2(\Psi^{(k)}, \Phi_s)$$
$$= -\sum_{s=1}^{m} (\Psi^{(k)} \Phi_s^T \Psi^{(k)} - \Phi_s) \quad (13)$$

As demonstrated in [33], $\Delta\Psi^{(k+1)}$ (red triangle on tangent plane $\mathcal{T}_{\Psi^{(k)}}\mathcal{M}_H$ in **Fig. 2**) is the updated position of the estimated mean and the red arrow specifies the direction from the prior estimation $\Psi^{(k)}$ to the new latent mean on the manifold.

(2) We map the mean tangent $\Delta\Psi^{(k+1)}$ back to the Stiefel manifold to obtain the new estimation of the manifold center $\Psi^{(k+1)} = exp_{\Psi^{(k)}}(\Delta\Psi^{(k+1)})$ (red circle in **Fig. 2**) by Eq. 5-6.

By iteratively calculating the optimal descent direction and mapping it back to the Stiefel manifold, we can obtain the optimal manifold mean $\Psi$, *i.e.,* the common harmonic waves. The entire optimization scheme is summarized in **Table II**.

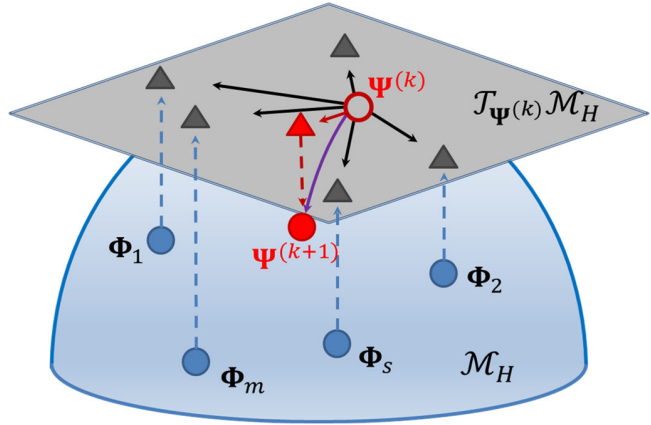

**Fig. 2** Illustration for the optimization of common harmonic waves on the Stiefel manifold. Individual harmonic waves (blue solid circle) are located on the Stiefel manifold $\mathcal{M}_H$ (blue hemisphere). The individual harmonic waves are projected to corresponding point (grey triangle) in the tangent space $\mathcal{T}_{\Psi^{(k)}}\mathcal{M}_H$ (grey flat plane) of $k^{th}$ manifold mean $\Psi^{(k)}$ (red hollow circle). The mean tangent (red triangle) is calculated based on all gradient directions (black arrow). Finally, the new manifold mean $\Psi^{(k+1)}$ (red solid circle) is estimated by mapping the mean tangent back to the Stiefel manifold. The manifold trajectory from $\Psi^{(k)}$ to $\Psi^{(k+1)}$ is depicted by purple arrow.

**Discussion.** Regarding the dimension reduction, we determine $p$ based on the distribution reconstruction error between the observed Laplacian matrix and the reconstructed Laplacian matrix using only the top $p$ smallest Eigen-values and Eigen-vectors. Empirically, we select $p$ around the tipping point that the decrease of reconstruction error is only marginal as $p$ increases. As shown in **Table II**, our optimization scheme consists of two subproblems. The proof of solving Eq. 9 using GPI is supported by the theorem 1-2 in [32]. The Weiszfeld algorithm [33] has been used in many computer vision applications with proof of convergence.

Table II Algorithm for common harmonics detection

| | |
|---|---|
| Parameters: $\lambda; \varepsilon_1; \varepsilon_2; \gamma;$ | |
| **Input:** | Adjacency matrix $W_s \in \mathbb{R}^{n \times n}$, $s = 1,2,\cdots,m$ |
| **Init.** | Calculate Laplacian matrix $L_s = D_s - W_s$, where $d_i = \sum_{j=1}^{n} W_{ij}$ ; |
| | Calculate positive definite matrix $\tilde{L}_s = \beta I - L_s$. Set $\beta$ be the dominant Eigen-value of $L_s$ ; |
| | Initialize common network harmonic waves $\Psi = eig(\frac{1}{m}\sum_{s=1}^{m} L_s) \in \mathbb{R}^{n \times p}$; |
| | Initialize orthogonal matrix $\Phi_s \in \mathbb{R}^{n \times p}$ through the Eigen-decomposition of Laplacian matrix $L_s$; |
| | Initialize parameter $\lambda = 0.01$, $\gamma = 0.01$; |
| **do** | |
| | **For** $s = 1,2,\cdots, m$ **do** |
| |    **do** |
| |       Update $\Theta = \tilde{L}_s \Phi_s^{(k)} + \lambda \Psi$ ; |
| |       Compute $U\Sigma V^T = \Theta$ via the compact SVD method of $\Theta$ where $U \in \mathbb{R}^{n \times p}$, $\Sigma \in \mathbb{R}^{p \times p}$ and $V \in \mathbb{R}^{p \times p}$; |
| |       Update $\Phi_s^{(k+1)} = UV^T$; |
| |    **While** $\left\|\Phi_s^{(k+1)} - \Phi_s^{(k)}\right\| < \varepsilon_1$ |
| | **End for** |
| | Set start point $\Psi^{(1)} = \Phi_1$ ; |
| | **do** |
| |    $\Delta \Psi^{(k+1)} = -\lambda \sum_{s=1}^{m}(\Psi^{(k)} \Phi_s^T \Psi^{(k)} - \Phi_s)$ ; |
| |    $\Psi^{(k+1)} = exp_{\Psi^{(k)}}(\gamma \Delta \Psi^{(k+1)})$ ; |
| | **While** $\|\Delta \Psi\| < \varepsilon_2$ |
| | Update $\Psi = \Psi^{(k+1)}$ ; |
| | Compute $New_{cost} = \sum_{s=1}^{m}\{tr(\Phi_s^T L_s \Phi_s) + \lambda d^2(\Psi, \Phi_s)\}$; |
| | $\varepsilon = abs(New_{cost} - Old_{cost})$; |
| | Update $Old_{cost} = New_{cost}$; |
| **While** | $\varepsilon$ is less than a pre-defined threshold. |
| **Output** | Common network harmonic waves $\Psi$. |

## 2.5 Application in Network Neuroscience

Advanced neuroimaging technology such as MRI and diffusion-weighted MRI allows us to study white matter fiber tracks associated with the progression of cognitive decline. Mounting evidence shows that neurodegenerative diseases such as Alzheimer's disease (AD) can be understood as a disconnection syndrome where the large-scale brain network is progressively disrupted by neuropathological processes [3]. Our proposed harmonic based network analysis approach provides a new methodology to analyze these spatio-temporal changes of the neuro-pathology in the progression of AD.

**Image processing.** As shown in **Fig. 3**, we parcellate the cortical surface into 148 cortical regions based T1-weighted MR image [34] and then apply surface seed-based probabilistic fiber tractography [34] using the diffusion tensor imaging (DTI) data, thus producing a $148 \times 148$ anatomical connectivity matrix. Note, the weight of the anatomical connectivity is defined by the number of fibers linking two brain regions normalized by the total number of fibers in the whole brain. Furthermore, we calculate the mean cortical thickness as well as the standard uptake value ratio (SUVR) of the amyloid deposition for each brain region and then assemble them into a column vector, denoted by $\boldsymbol{f}^s$.

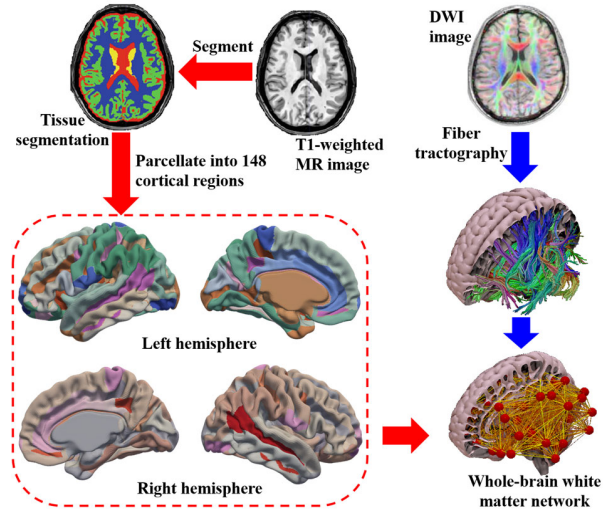

**Fig. 3** Image processing pipeline to construct structural brain network from diffusion-weighted MR images.

**Harmonic analyses.** For each common harmonic wave $\boldsymbol{\psi}_h$ ($h^{th}$ column vector in $\boldsymbol{\Psi}$), we first calculate the harmonic power coefficient of $\boldsymbol{\psi}_h$ to the observed cortical thickness (or amyloid deposition) vector $\boldsymbol{f}^s$ of $s^{th}$ individual subject by:

$$\alpha_h^s = \langle f^s, \psi_h \rangle \tag{14}$$

Furthermore, we can calculate the corresponding harmonic-specific energy of $\psi_h$ by:

$$E_h^s = |a_h^s|^2 \tag{15}$$

The total harmonic energy of brain injury (manifested by cortical thickness or amyloid deposition) with respect to the harmonic waves $\Psi_h$ is measured by:

$$E_{total}^s = \sum_{h=1}^{p} E_h^s \tag{16}$$

## 3. Results

To evaluate the power of our new network harmonic analysis approach, we compare the performance of using the common harmonics $\Psi$ optimized by our manifold learning method to two alternative sets of harmonic waves: (1) $\bar{\Psi}$ by simple averaging individual Eigen-systems, and (2) $\tilde{\Psi}$ by first averaging the adjacency matrices and then applying SVD to the average adjacency matrix. In the following, we call $\bar{\Psi}$, $\tilde{\Psi}$, and $\Psi$ as the arithmetic mean, the pseudo manifold mean (aka. pseudo common harmonics), and Stiefel manifold mean (aka. our common harmonics), respectively.

### 3.1 Experiments on Synthetic Data

Here, we synthesize a set of 3D rotation (orthonormal) matrices, which are represented as the unit quaternions. A quaternion is denoted as $q = (a, v)$, where $a$ is the real quantity and $v = bi + cj + dk$ with three imaginary quantities $(b, c, d)$. Let $e = (1, 0)$ be the identity quaternion. The transformation between quaternion and rotation matrix can be represented as

$$R = \begin{bmatrix} 1 - 2c^2 - 2d^2 & 2bc - 2ad & 2ac + 2bd \\ 2bc + 2ad & 1 - 2b^2 - 2d^2 & 2cd - 2ad \\ 2bd - 2ac & 2ab + 2cd & 1 - 2b^2 - 2c^2 \end{bmatrix} \tag{17}$$

where $a = \cos\left(\frac{1}{2}\theta\right)$, $b = \sin\left(\frac{1}{2}\theta\right)u_x$, $c = \sin\left(\frac{1}{2}\theta\right)u_y$, $d = \sin\left(\frac{1}{2}\theta\right)u_z$, as well as $\theta$ and $\boldsymbol{u} = (u_x, u_y, u_z)$ denote rotation angle and rotation axis, respectively.

We generate a random collection of twenty quaternions as follows. First, we set the quaternion with no rotation as the ground truth (starting point), which is displayed in green in **Fig. 4**. Second, given rotation axis $\boldsymbol{u}$, the rotation angles are sampled from a zero-mean Gaussian distribution with a standard deviation $\sigma = \pi/15$. Third, twenty rotation matrices are obtained through Eq. 17, centered on the identity matrix. Among them, ten rotation matrices are shown in the first two rows of **Fig. 4**.

Since we do not have the adjacency matrices, we apply naïve averaging and our Stiefel manifold learning method to estimate the common quaternion from the 20 random perturbative quaternions. The arithmetic mean $\overline{\boldsymbol{\Psi}}$ and our Stiefel manifold mean $\boldsymbol{\Psi}$ are shown in **Fig. 4(b)** and **(c)**, respectively. It is clear that (1) The Stiefel mean is very close to the ground truth on the manifold; (2) The arithmetic mean is out of the manifold surface (non-orthogonal matrix), as the three rotation axes are not perpendicular to each other; (3) Our iterative manifold optimization can quickly converge to the latent manifold mean, as indicated by the red trajectory in **Fig. 4**. Although we initialize our optimization from a single individual quaternion (#10) in **Fig. 4**, no significant difference has been found across the Stiefel mean results initialized with different individual quaternions. We further examine the replicability in Section 3.2.2.

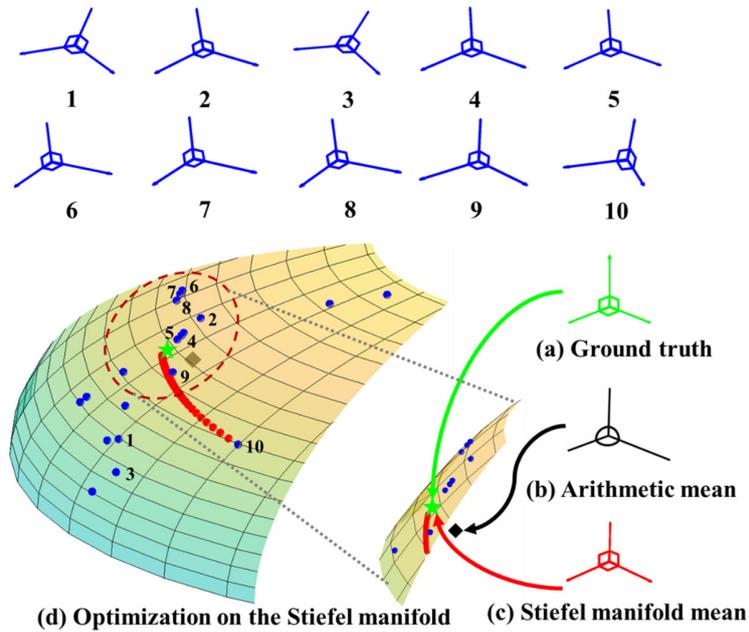

**Fig. 4** Comparison of arithmetic mean and our Stiefel manifold mean across individual orthonormal matrices. Top: 10 examples of matrices (first two rows) generated by different rotation of identity matrix (a). Bottom: the ground truth (a), arithmetic mean (b), and our Stiefel manifold mean (c) on the Stiefel manifold (d). It is clear that the manifold center estimated by our manifold optimization method is more reasonable than the arithmetic mean which uses simple averaging operation defined in the Euclidean space.

## 3.2 Experiments on Real Data of Alzheimer's Disease

### 3.2.1 Description of datasets and experiment setup

**Training data for learning common harmonic waves.** In total 94 subjects were selected from ADNI database to learn the common harmonic waves, which consists of 17 Cognitive Normal (CN), 18 Significant Memory Concern (SMC), 24 Early Mild Cognitive Impairment (EMCI), 16 Late Mild Cognitive Impairment (LMCI), and 19 Alzheimer's Disease (AD). Each subject has both T1-weighted MRI and diffusion-weighted MRI scans. The demographic information is shown in **Table III**. Following the image processing pipeline in **Fig. 3**, we constructed the structural network for each subject which consisted of 148 nodes.

Table III Demographic information of training data in ADNI database

| Gender | Number | Range of Age | Average Age | CN | SMC | EMCI | LMCI | AD |
|---|---|---|---|---|---|---|---|---|
| Male | 47 | 55.0~90.3 | 74.3 | 7 | 5 | 16 | 8 | 11 |
| Female | 47 | 55.6~87.8 | 73.0 | 10 | 13 | 8 | 8 | 8 |
| **Total** | **94** | **55.0~90.3** | **73.7** | **17** | **18** | **24** | **16** | **19** |

**Testing data for identifying frequency-based alterations in AD.** In addition, we selected another 50 CN subjects and 47 AD subjects from ADNI data as the testing data. As was done with the training data, the cortical surface of each subject was parcellated into 148 regions, and the mean cortical thickness and the standard uptake value ratio (SUVR) of amyloid deposition for each region were computed.

**Experiment setup on real AD network data.** In the following experiments, we only compare the performance between the pseudo manifold mean $\widetilde{\Psi}$ and the Stiefel manifold mean $\Psi$, as the non-orthogonality of the arithmetic mean $\overline{\Psi}$ (**Fig. 4(b)**) is not ideal for explaining frequency-based alterations. The number of harmonic waves $p$ is set to 60. First, we evaluate the replicability of the common harmonic waves using our proposed learning-based method via a resampling test in Section 3.2.2. Next, we investigate whether the oscillation patterns in the common harmonic waves underline the neurodegenerative process in Section 3.2.3. This sets the stage for applying the learned common harmonic waves to identify harmonic-based alterations in the cortical thickness (Section 3.2.4) and amyloid level data (Section 3.2.5).

**3.2.2 Evaluation of the replicability**

In this experiment, we evaluate the replicability of the learned common harmonic waves via resampling tests. Specifically, we apply the following resample procedure to generate 50 test/retest datasets from the training data: (1) randomly sample 70 networks from the 94 training network data; (2) continue to sample another two sets of networks from the remaining 24 subjects separately, each with 5 networks; (3) form two paired cohorts by combining the networks sampled in step 1 and 2. Then, we deploy our Stiefel manifold learning method on two datasets independently. Because two paired cohorts only have 6.7% (5/75) differences in terms of network data, we can evaluate the replicability of our method by examining whether there exists a significant difference at each element in the harmonic waves via the paired $t$-test. Fewer elements showing significance indicates better replicability. Since each row in the harmonic matrix is associated with one brain region, we can map the significant findings ($p < 0.01$) onto the cortical surface in **Fig. 5(a)** and **(b)**. It is apparent that our Stiefel manifold learning method yields more consistent (more replicable) common harmonic waves across the test/retest datasets in the resampling test.

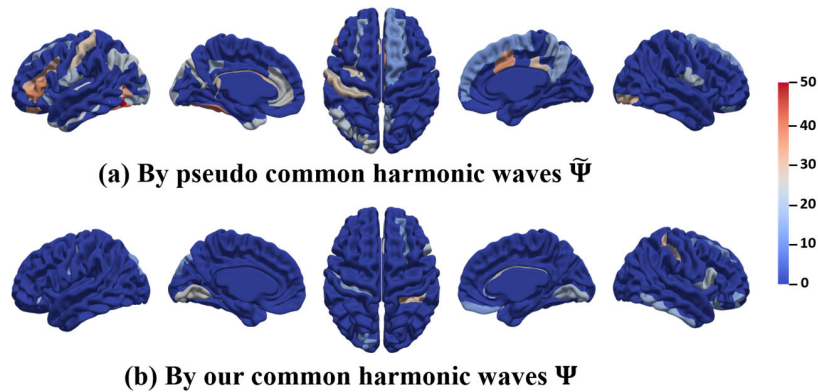

(a) By pseudo common harmonic waves $\widetilde{\Psi}$

(b) By our common harmonic waves $\Psi$

**Fig. 5** The replicability test results of the pseudo common harmonic waves $\widetilde{\Psi}$ (a) and our common harmonic waves $\Psi$ (b), where the color on the cortical surface reflects the number of times with failed replicability tests. Our methods show higher replicability compared to the pseudo harmonic waves.

**3.2.3 Association between the Oscillation Patterns in Harmonic Waves and Neurodegenerative Process**

In this experiment, we are investigating whether the learned common harmonic waves capture information related to the neurodegenerative process in AD. After we compute the common harmonic waves on the training data, we repeat the following steps on the testing data with 50 replicates: (1) randomly select 30 out of 47 AD subjects and 30 out of 50 CN subjects as training data and form the amyloid vector $\boldsymbol{f}$; (2) identify the harmonic power difference $\alpha_h$ between CN and AD for each harmonic wave $\boldsymbol{\psi}_h$; (3) calculate positive power $\alpha_h^+ = \langle \boldsymbol{f}, \boldsymbol{\psi}_h^+ \rangle$ and negative power $\alpha_h^- = |\langle \boldsymbol{f}, \boldsymbol{\psi}_h^- \rangle|$ of the remaining 37 subjects (testing data), where $\psi_h^+$ and $\psi_h^-$ present the positive-only and negative-only segments in each $\psi_h$; (4) apply the *t*-test to detect the statistical CN vs AD difference of $|\alpha_h^+ - \alpha_h^-|$ on each harmonic identified in (2).

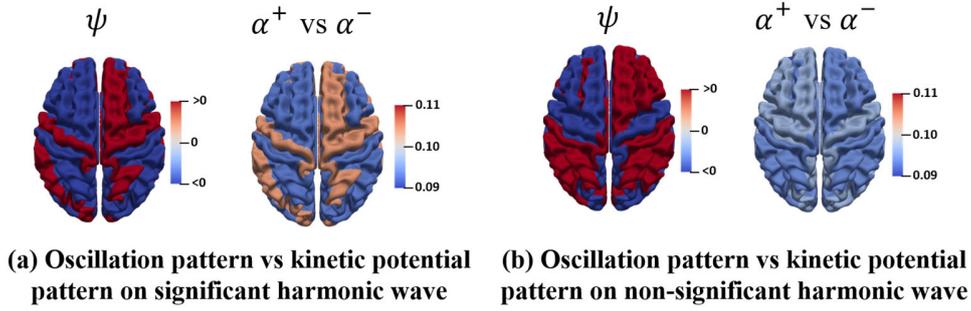

(a) Oscillation pattern vs kinetic potential pattern on significant harmonic wave

(b) Oscillation pattern vs kinetic potential pattern on non-significant harmonic wave

**Fig. 6** The replicability test results of the pseudo common harmonic waves $\widetilde{\boldsymbol{\Psi}}$ (a) and our common harmonic waves $\boldsymbol{\Psi}$ (b), where the color on the cortical surface reflects the number of times with failed replicability tests. Our methods show higher replicability compared to the pseudo harmonic waves.

First, we find that on average 16.6% (9.95/60) pseudo common harmonic waves and 13.8% (8.28/60) our common harmonic waves show significant harmonic power differences in the amyloid data in step (2). The detailed statistics are shown in **Table IV**. Next, we test the hypothesis that the positive-negative harmonic power difference (kinetic potentials of amyloid level due to the oscillations in harmonic waves) between $a_h^+$ and $a_h^-$ is the factor leading to such significance. The rationale is that the oscillation patterns in the harmonic waves are correlated with the observations of pathological burdens if not only (1) the harmonic power shows significant differences between CN and AD, but also (2) the positive-negative power differences also manifest significant differences between CN and AD. As such, the observed harmonic waves can serve as biological indicators (factors) for the progression of AD. As a piece of evidence for the above hypothesis, we display the oscillation mapping of both a significant and a non-significant harmonic wave on the cortical surface in **Fig. 6(a)** left and **Fig. 6(b)** left, respectively. This visualization also shows the associated cortical mapping of elementwise vector multiplication between $\boldsymbol{\psi}_h$ and $\boldsymbol{f}$ at the right side of **Fig. 6(a)-(b)**. It is apparent that the two cortical mappings for the significant harmonic wave in **Fig. 6(a)** have a strong resemblance, which is also supported by the statistical significance between $\alpha^+$ and $a^-$ ($p < 10^{-5}$). On the contrary, such resemblance is not presented in the non-significant harmonic wave (**Fig. 6(b)**), where no significance has been detected between $\alpha^+$ and $a^-$ ($p = 0.27$).

**Table IV** Statistics of kinetic potentials

| Methods | Har. Power | +/- Power |
|---|---|---|
| Pseudo common harmonics $\widetilde{\Psi}$ | 9.95 ± 2.28 | 3.48 ± 1.62 |
| Our common harmonics $\Psi$ | 8.28 ± 1.47 | 5.23 ± 1.17 |

As shown in the last column in **Table IV**, 63% (5.23/8.28) of the identified significant common harmonic waves in $\Psi$ support such a hypothesis since average 5.23 common harmonic waves exhibit the statistical significance of $|\alpha_h^+ - \alpha_h^-|$ ($p < 10^{-3}$) in step (4). As a comparison, we find only less than 35% (3.48/9.95) of the harmonics in $\widetilde{\Psi}$ show CN vs AD significance in both harmonic power $\alpha_h$ and positive-negative power difference $|\alpha_h^+ - \alpha_h^-|$. The results in **Table IV** indicate that the oscillation patterns in our learned common harmonic waves have more statistical correlations with pathological neurodegeneration events. In the following two experiments, we apply our learned common harmonic waves $\Psi$ to identify frequency-based harmonic alterations in the context of neurodegeneration biomarker measured by the cortical thickness (3.2.4) and amyloid deposition (3.2.5).

**3.2.4 Identifying frequency-based harmonic alterations in cortical thickness**

A plethora of neuroimaging studies found morphometry differences between CN and AD cohorts. Since the common harmonic waves discovered by our manifold learning method are potentially related to the neurodegenerative process as demonstrated in Section 3.2.3, we explore the frequency-based alterations of cortical thickness values from the testing data that are relevant to AD progression by using the learned common harmonics from the training data. First, we measure the total harmonic energy of cortical thickness for each subject and plot the statistics (mean and standard deviation) for CN and AD groups separately in **Fig. 7(a)**, where the AD group (15.9 ± 4.6) holds significantly lower ($p < 10^{-4}$) total energy than CN group (19.6 ± 4.4). Furthermore, we plot the distribution of total energy in **Fig. 7(b)**. These results support the evidence that neurodegeneration in AD subjects is associated with reduced neuroanatomical structural integrity. Second, we examine the cross-sectional energy difference for each harmonic, where the mean harmonic-specific energy for CN and AD are shown in the outer and inner rings in **Fig. 7(c)**. In addition, the Fisher score $J_F$ (the ratio between inter-class mean and intra-class variance) of the harmonic-specific energy between CN and AD subjects is shown in the outermost ring in **Fig. 7(c)**, where the harmonic waves exhibiting significant energy differences are tagged with a red star '*'. The CN-to-AD difference magnitude at each harmonic wave is displayed in **Fig. 7(d)**. These significant harmonic waves may be critically important in determining the propagation of neuropathological burdens across the brain networks.

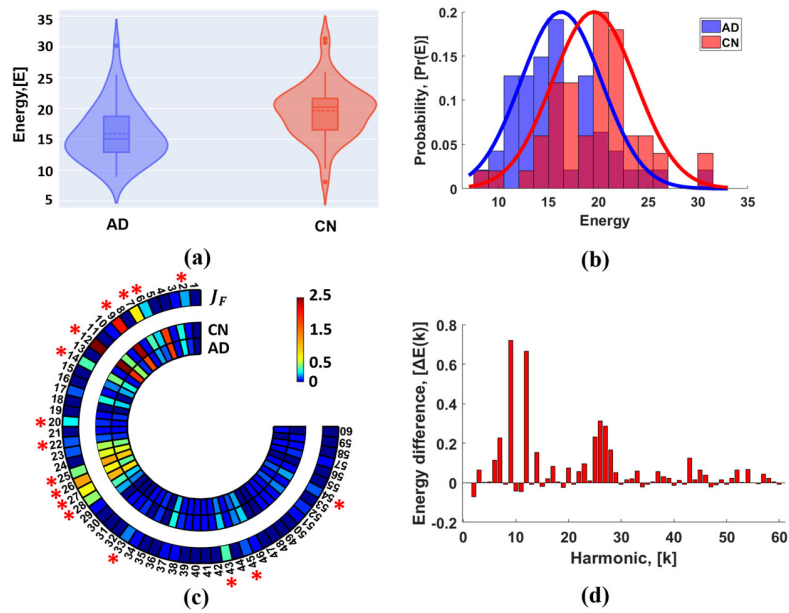

**Fig. 7** Harmonic alterations between CN and AD identified using the learned common harmonics **Ψ** on cortical thickness. (a)-(b): Significant difference of total harmonic energy has been detected between CN and AD. (c): Harmonic waves exhibiting significant energy difference between CN and AD cohorts. (d): The plot of CN-to-AD energy difference of each harmonic wave.

### 3.2.5 Identifying frequency-based harmonic alterations in amyloid deposition

Similarly, we calculate the total harmonic energy of the amyloid deposition for each subject and plot the results in **Fig. 8(a)** and **(b)**, where the AD group $(4.41 \pm 1.85)$ has significantly higher $(p < 10^{-4})$ total energy than CN group $(3.19 \pm 1.40)$. In addition, we show the statistical significance in energy difference and the CN-to-AD energy difference magnitude for each harmonic wave in **Fig. 8(c)** and **(d)**, where there are a total of 15 harmonic waves exhibiting significant difference $(p < 0.01)$ between CN and AD, in terms of harmonic energy of amyloid deposition. These results suggest that the aggregation of amyloid peptides is associated with topological features of the brain networks that underlie the network harmonics.

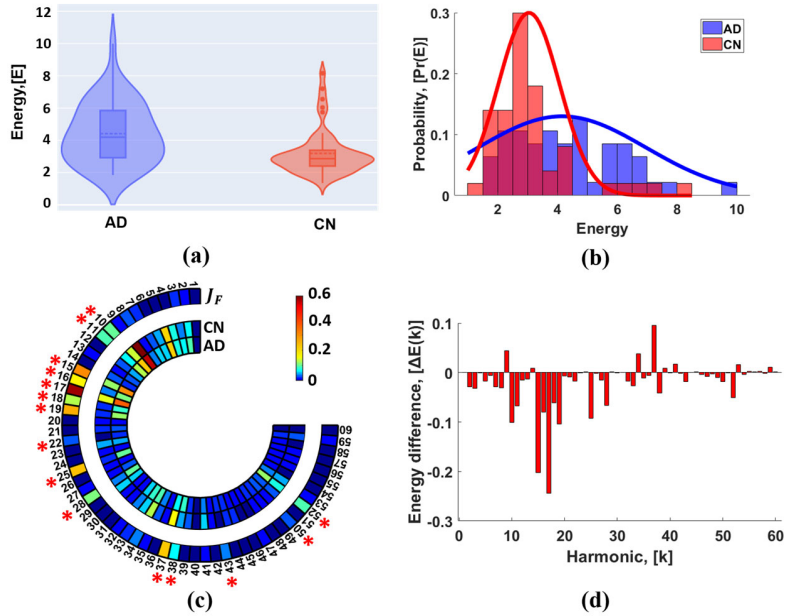

**Fig. 8** Harmonic alterations between CN and AD identified using the learned common harmonics $\Psi$ on amyloid deposition. (a)-(b): Significant difference of total harmonic energy has been detected between CN and AD. (c): Harmonic waves exhibiting significant energy difference between CN and AD cohorts. (d): The plot of CN-to-AD energy difference of each harmonic wave.

### 3.2.6 Discussions

The deposition of Amyloid plaques is one of the hallmarks of AD. Both human and animal data suggest a causal upstream role for amyloid-β in the pathogenesis of AD, which may be sufficient to cause downstream pathologic changes leading to cognitive decline [35]. Our finding of frequency-based harmonic alterations in amyloid deposition complements the current neuroscience and clinical literature, with the AD population having greater amyloid harmonic energy than the CN group (**Fig. 8(d)**). Similarly, reductions in cortical thickness are thought to reflect neuro-degeneration associated with AD progression. As shown in **Fig 7(d)**, CN subjects have more cortical thickness harmonic energy than AD subjects in most of the harmonic frequency bands, which indicates that degeneration (structural atrophy) is more profound in the AD than the CN cohort.

Furthermore, we found 16 harmonic waves for cortical thickness and 15 out of 60 common harmonic waves for amyloid that were significant differences between CN and AD. We display the oscillation pattern of the identified harmonic waves for neuro-degeneration (cortical thickness) and amyloid deposition in **Fig. 9**, where the shared harmonic waves by the cortical thickness and amyloid deposition are shown at the top. In addition, the top 4 significant harmonic waves with the smallest $p$-value specific to cortical thickness and amyloid burden are shown in middle and bottom in **Fig. 9**, respectively.

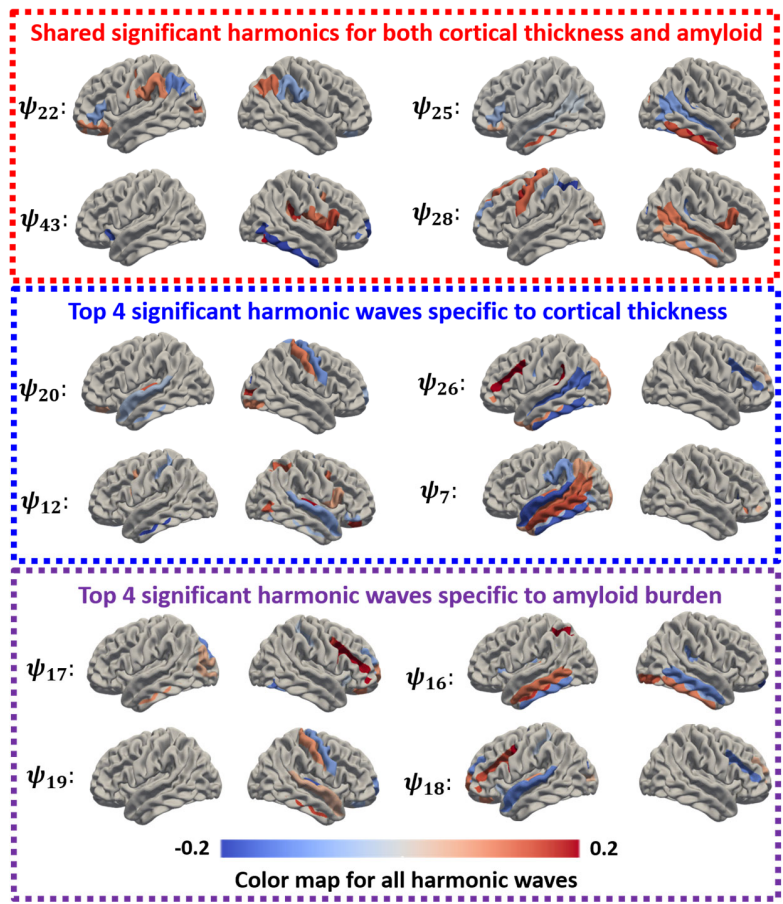

**Fig. 9** The spatial pattern of significant harmonic waves in the context of cortical thickness (blue box), and amyloid deposition (purple box). Their shared harmonic waves are displayed in the red box.

Different neurodegenerative diseases exhibit distinct network alteration patterns [17]. For example, AD is associated with atrophy and hypometabolism in the posterior hippocampal, cingulate, temporal, and parietal regions, which collectively resemble the default mode network (DMN) [36, 37]. In contrast to AD, behavior variant frontotemporal dementia (bvFTD) preferentially affects the salience network (SN) [17, 36]. Here, we examine the association between the oscillation pattern and these large-scale networks. *First*, we mark the location of the top ten crossing-zeros in each harmonic wave which has the largest difference magnitude. In general, 22-24% of the crossing-zeros are found falling in the DMN, compared to only 2% of them are associated with the SN. *Second*, we calculate the frequency of each node being touched by the crossing-zeros across all significant harmonic waves. We show the node frequency maps by cortical thickness and amyloid in the middle and right of **Fig. 10**, respectively. It is clear that much more crossing-zeros are associated with DMN (top) than SN (bottom), which is aligned with the current findings in AD.

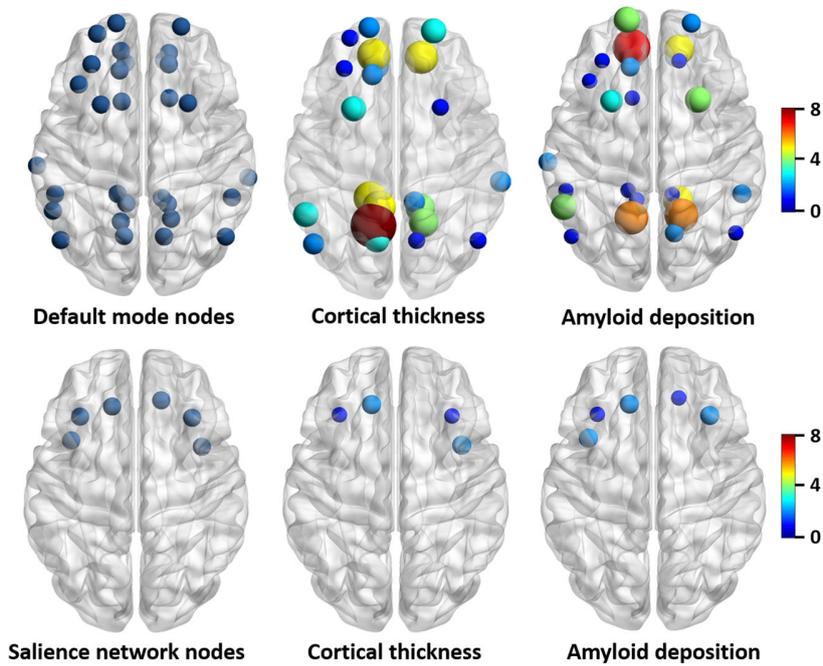

**Fig. 10** The spatial alignment of crossing-zeros in the identified significant common harmonic waves with respect to default mode network and salience network. Left: nodes belong to DMN and salience network. Middle: in the context of cortical thickness. Right: in the context of amyloid deposition.

## 4. Conclusion

In this paper, we present a new network harmonic analysis approach that offers a new window into the investigation of frequency-based alterations between different clinical and research study populations. To achieve this, we propose a manifold optimization method to find the set of common harmonic waves from the native Eigen-systems of individual brain networks. The resulting shared reference space spanned by the common harmonic waves allows us to quantify the individual kinetic differences in terms of propagating neuro-pathological events across brain networks. We have evaluated the power of the common harmonic waves in discovering harmonic-specific alterations between CN and AD. More consistent and reasonable results were achieved by our manifold learning method, compared to the current methods which use Euclidean operations on the manifold data.

In the future, we plan to apply our new network harmonic analysis approach to other neurological disorders which manifest network dysfunction syndrome such as frontotemporal dementia and schizophrenia.